\documentclass[a4paper,10pt,twoside]{cpc-hepnp}

\usepackage{multicol}
\usepackage{graphicx}
\usepackage{booktabs}
\usepackage{amssymb,bm,mathrsfs,bbm,amscd}
\usepackage[tbtags]{amsmath}
\usepackage{lastpage}

\begin{document}

\fancyhead[R]{\footnotesize Submitted to 'Chinese Physics C'}

\footnotetext[0]{Received 21 June 2013}

\title{Physical design of a wavelength tunable fully coherent VUV source using self-seeding free electron laser\thanks{Supported by Major State Basic Research Development Program of China ( Grant No. 2011CB808301) and National Natural Science Foundation of China (Grant No: 11205156) }}

\author{%
      LI He-Ting
\quad JIA Qi-Ka} \maketitle

\address{%
National Synchrotron Radiation Laboratory, University of Science and
Technology of China, Hefei, 230029, Anhui, China\\
}

\begin{abstract}
In order to meet requirements of the synchrotron radiation users, a fully coherent VUV free electron laser (FEL) has been preliminarily designed. One important goal of this design is that the radiation wavelength can be easily tuned in a broad range (70-170 nm). In the light of the users' demand and our actual conditions, the self-seeding scheme is adopted for this proposal. Firstly, we attempt fixing the electron energy and only changing the undulator gap to varying the radiation wavelength, but the analysis implies that it is difficult because of the great difference of the power gain length and FEL efficiency at different wavelength. Therefore, dividing the wavelength range into three subareas is considered. In each subarea, a constant electron energy is used and the wavelength tuning is realized only by adjusting the undulator gap. The simulation results shows that this scheme has an acceptable performance.
\end{abstract}



\footnotetext[0]{\hspace*{-3mm}\raisebox{0.3ex}{$\scriptstyle\copyright$}2013
Chinese Physical Society and the Institute of High Energy Physics
of the Chinese Academy of Sciences and the Institute
of Modern Physics of the Chinese Academy of Sciences and IOP Publishing Ltd}%

\begin{multicols}{2}

\section{Introduction}

Drived by scientific research demands of the synchrotron radiation users at Hefei Light Source (HLS), a 70-170 nm wavelength tunable fully coherent free electron laser is desired. It is a very broad range of wavelength for an FEL facility. In addition, it is also expected that the radiation intensity gradually increases with the wavelength tuning from short to long.
Last year, a similar source has been projected, named Dalian Coherent Light Source (DCLS) [1], which applies the high-gain harmonic generation (HGHG) [2] scheme with an optical parametric amplification (OPA) seed laser and is dedicated at the spectral regime of 50-150nm. In the light of the users' requirements and our actual conditions, the self-seeding scheme is selected in this proposal as a preliminary reaserch.
The self-seeding FEL [3-8] is a promising approach to significantly narrow the self amplified spontaneous emission (SASE) [9] bandwidth. Generally, a self-seeding facility consists of two undulators separated by a photon monochromator and an electron bypass. The two undulators are resonant to the same radiation wavelength. The SASE radiation generated by the first undulator passes through the monochromator to create a  transform-limited pulse, which is used as a seed laser in the second undulator. The electron bypass normally is a four-dipole chicane. The dispersion of the chicane smear out the electron bunch microbunching produced in the first undulator. The monochromatized laser pulse is amplified by interacting with the non-bunched electron bunch in the second undulator. The required seed power for the second undulator must dominate over the shot noise power within the gain bandpass. Usually, the self-seeding FEL requires a longer undulator. However, it does not need the OPA seed laser.
In this article we preliminarily design the 70-170 nm FEL using the self-seeding scheme. Particular emphasis is laid on the FEL physics. The wavelength tuning is implemented mainly by adjusting the undulator gap but not changing the electron energy. The reason for this is that, on one hand, as the electron energy is not very high here, the undulator natural focusing is considerable and the undulator length of self-seeding FEL is usually long, hence the focusing system and the bypass chicane both should be reset if the electron energy is changed. On the other hand, if tuning the radiation wavelength mainly by adjusting the electron energy, it will bring difficulties to the design of the monochromator and the user beam line because of the great difference of the radiation power at different wavelength. Therefore, we consider using a constant electron energy and only adjusting the gap in section 2, then in section 3 we divide the radiation wavelength range into three subareas and a constant electron energy is used in each one.

\section{Using one constant electron energy}

We select the NdFeB permanent magnet undulators with a remanence of~$\emph{B}_r=1.2$~. Its peak magnetic intensity B0 can be calculated by the empirical formula:
\begin{eqnarray}
B_0  = 3.495e^{ - 4.885\frac{g}{{\lambda _u }} + 1.41(\frac{g}{{\lambda _u }})^2 } ,{\rm{ }}0.07 \le \frac{g}{{\lambda _u }} \le 0.7
\end{eqnarray}
where ~$\emph{g}$~ and ~$\lambda _u$~  are the gap and period of the undulator respectively. For a planar undulator, the normalized magnetic parameter is,
\begin{eqnarray}
a_u  = 0.6605B_0 [{\rm{T}}]\lambda _u [{\rm{cm}}]
\end{eqnarray}
on the other hand, the two parameters are also constrained by the FEL resonance,
\begin{eqnarray}
\lambda _s  = \frac{{\lambda _u }}{{2\gamma ^2 }}(1 + a_u^2 )
\end{eqnarray}
where ~$\lambda _s$~ is the resonant wavelength and ~$\gamma$~  is the normalized electron energy.
\begin{center}
\includegraphics[width=8.0cm]{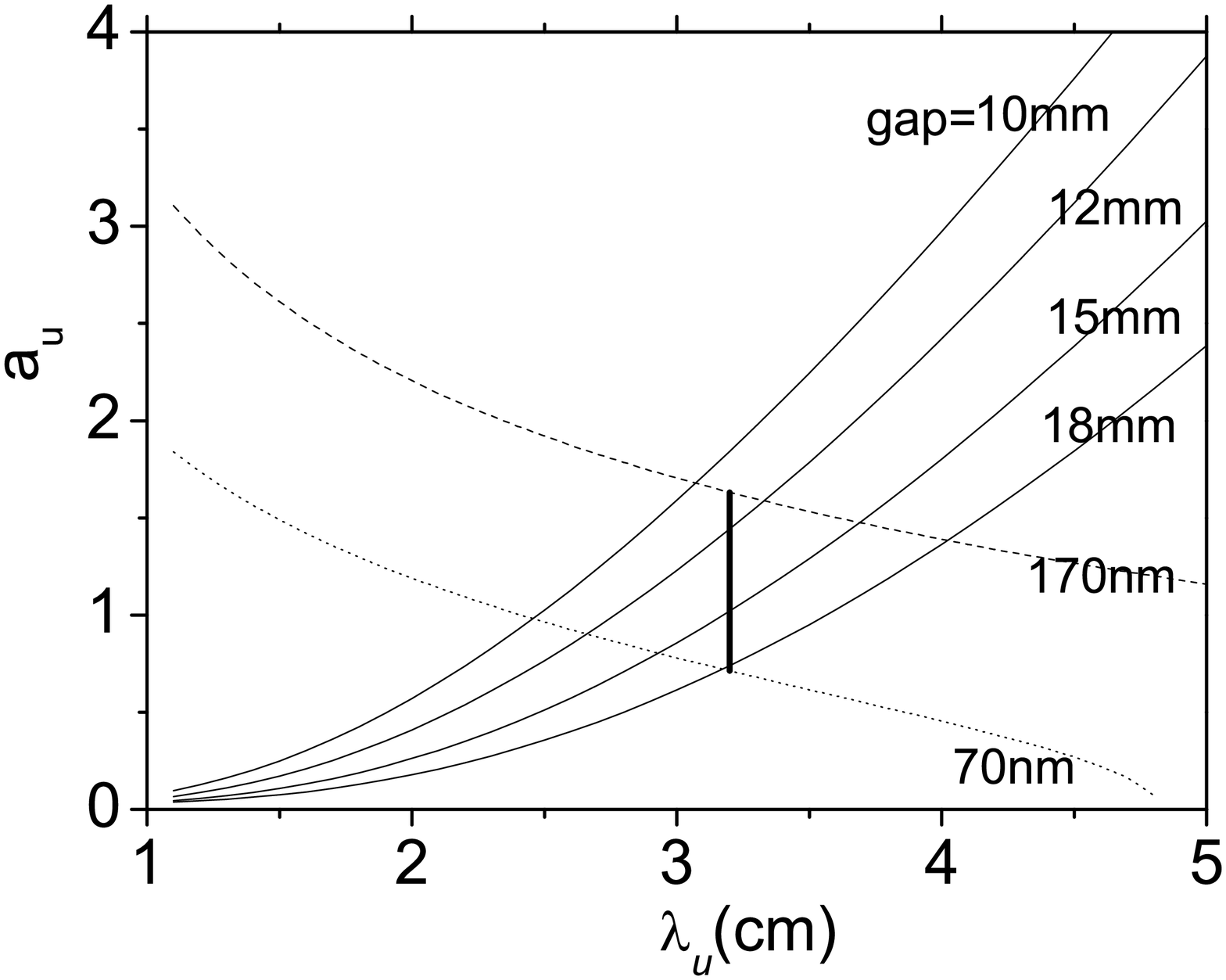}
\figcaption{\label{fig1} The undulator parameter ~$a_u$~ as functions of the undulator period ~$\lambda _u$~ for Formula (2) (line) and Formula (3) (dot line). The electron beam energy is assumed to 300 MeV.}
\end{center}
\begin{center}
\includegraphics[width=8.0cm]{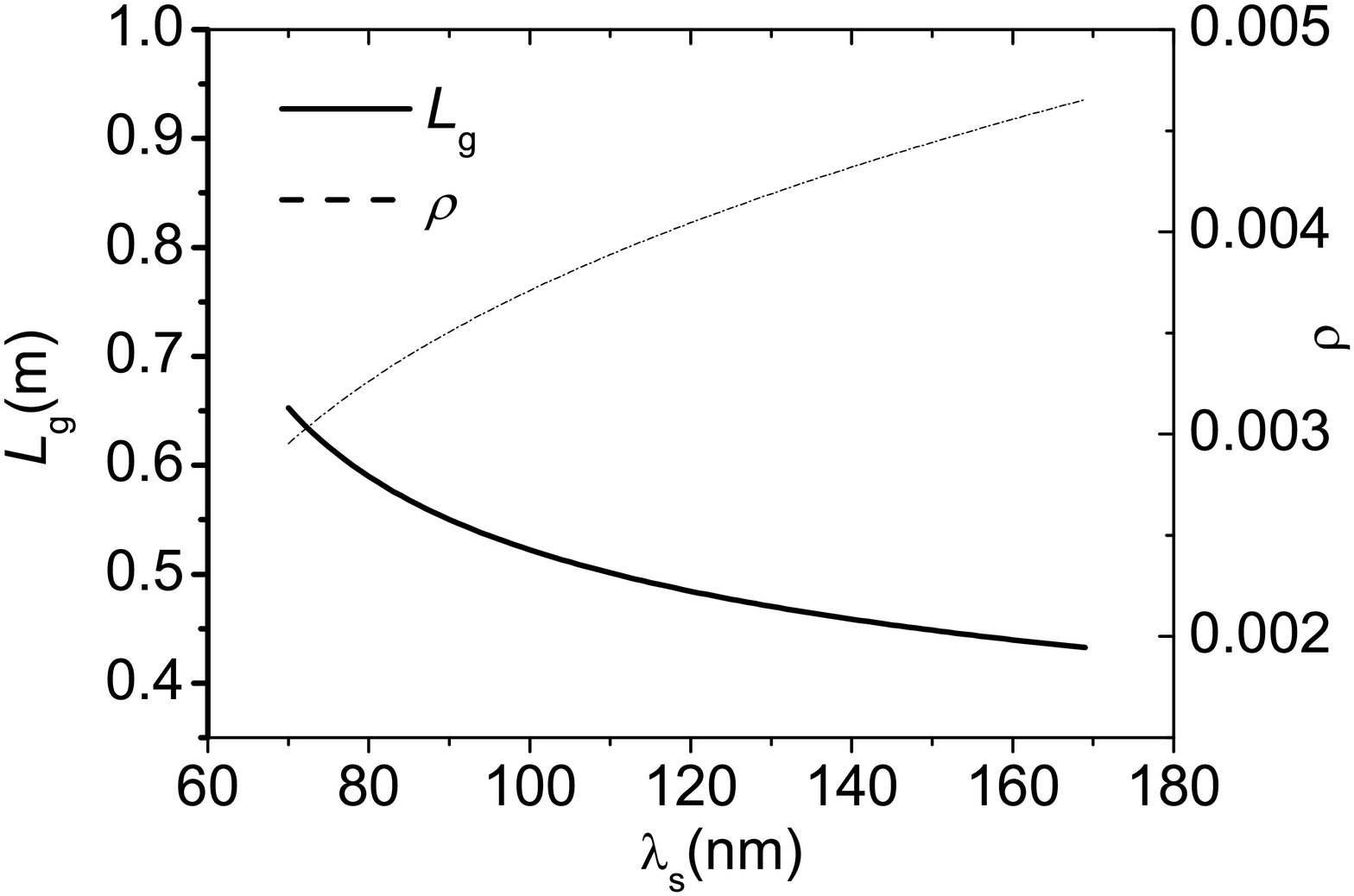}
\figcaption{\label{fig2} The FEL power gain length and pierce parameter varying with the radiation wavelength.}
\end{center}
\begin{center}
\includegraphics[width=8.0cm]{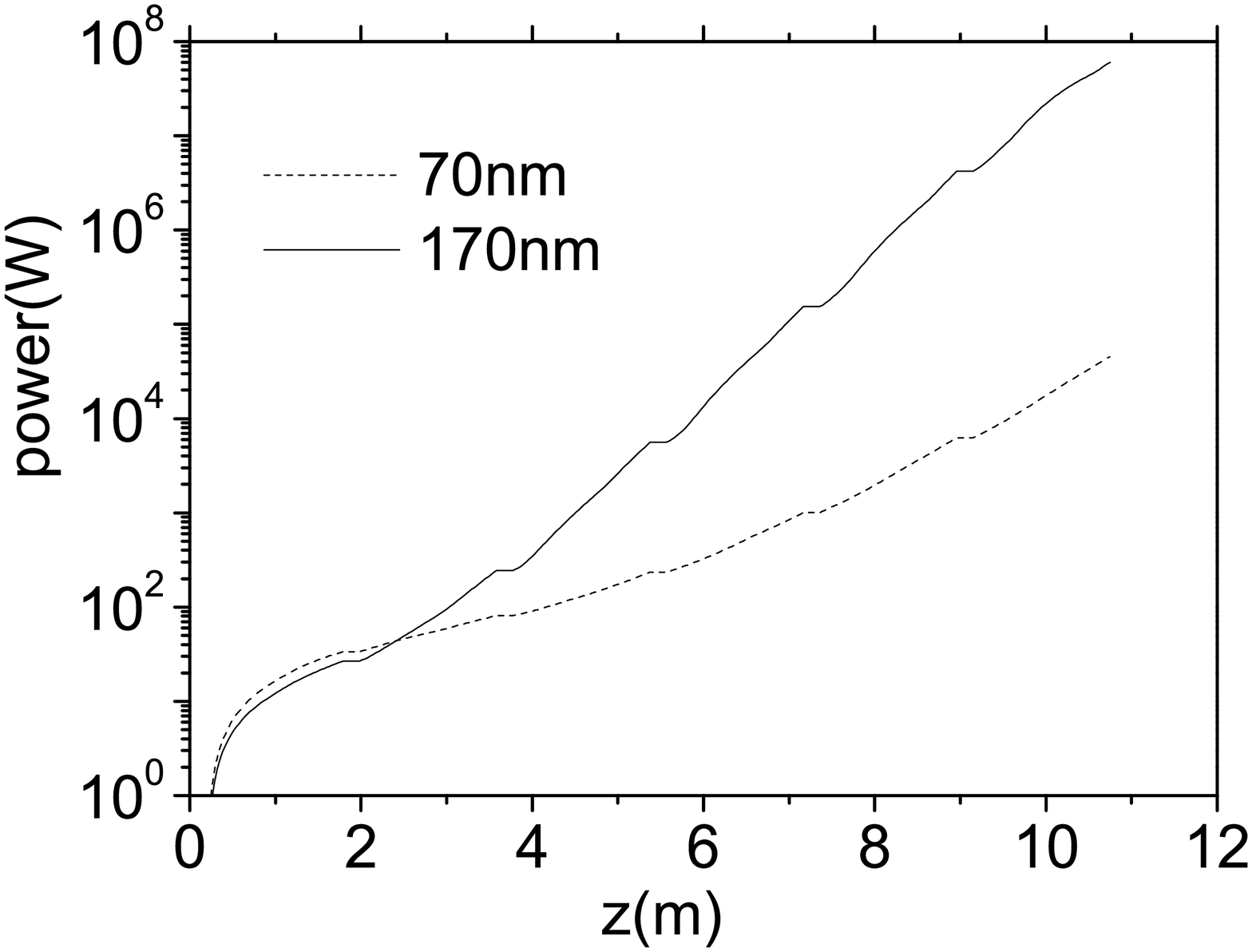}
\figcaption{\label{fig3} The FEL power of 70 nm and 170 nm in the first section.}
\end{center}

Figure 1 plots the curves of the undulator parameter au as functions of undulator period for Formula (2) and (3). Considering the construction technics of undulator, we choose the electron beam energy to be 300 MeV and the undulator period to be 3.2 cm. In this case,
when tuning the radiation wavelength from 70 nm to 170 nm, the undulator gap varies in a range of about 11.5-18 mm. The FEL power gain length and pierce parameter varying with the radiation wavelength are given in Fig.2. One can find that the power gain length decreases from 0.65 m to 0.35 m and the pierce parameter grows from 0.0028 to 0.0047 with radiation wavelength tuning from 70 nm to 170 nm. This may result in a problem that it is difficult to determine the length of the first section. If too short, the radiation intensity at the short wavelength will be not strong enough to seed the second section. Contrarily, if too long, the electron beam quality will degrade too much for the long wavelength because whose gain length is shorter. As Fig.3 shows, the 170 nm has a higher radiation power than 70 nm by three orders of magnitude around z=10 m.

Therefore, we consider to divide the whole wavelength range into several subareas.

\section{Dividing the wavelength range into three subareas}

We have investigated dividing the wavelength range into two subareas. The results imply that it can not solve the problem satisfactorily. Then three wavelength subareas is considered. In each subarea, the electron energy is constant and the focusing settings can be fixed. The electron energy optimization is similar as in Part 2. It has a upper limit corresponding to the minimum undulator gap. The detail is displayed in Table.1. The electron beam power is enhanced for the short wavelength so as to achieve a comparative power for all wavelength at the exit of the first section. The main parameters for our proposal is listed in Table.2. The simulation is based on the code GENESIS 1.3 [10]. It should be pointed out that we do not scan the initial  phase distribution of the electron beam as normal SASE simulations because in this paper we focus on how to reach the goal of broad wavelength range.

\begin{center}
\tabcaption{ \label{tab1}  The division of the wavelength range.}
\footnotesize
\begin{tabular*}{80mm}{c@{\extracolsep{\fill}}ccc}
\toprule $\lambda _s$ /nm & $\emph{E} _0$ /MeV   & Undulator gap /mm  \\
\hline
70-100\hphantom{00} & \hphantom{0}360 & 12-14.5 \\
100-130\hphantom{00} & \hphantom{0}315 & 12-13.75 \\
130-170\hphantom{0} & 275 & 12-13.8 \\
\bottomrule
\end{tabular*}
\end{center}

\begin{center}
\tabcaption{ \label{tab2}  Main parameters for this proposal.}
\footnotesize
\begin{tabular*}{80mm}{c@{\extracolsep{\fill}}ccc}
\toprule Parameter  & Specification  \\
\hline
$\emph{E} _0$ /MeV & $<360$ \\
Slice energy spread & $0.01\%$ \\
Peak current /A & 400  \\
Normalized emittance /mm・mrad & 1.3   \\
Bunch length (FWHM) /ps & 2.5  \\
undulator period /cm & 3.2 \\
undulator section length  /m & 1.6 \\
\bottomrule
\end{tabular*}
\end{center}
\begin{center}
\includegraphics[width=8.0cm]{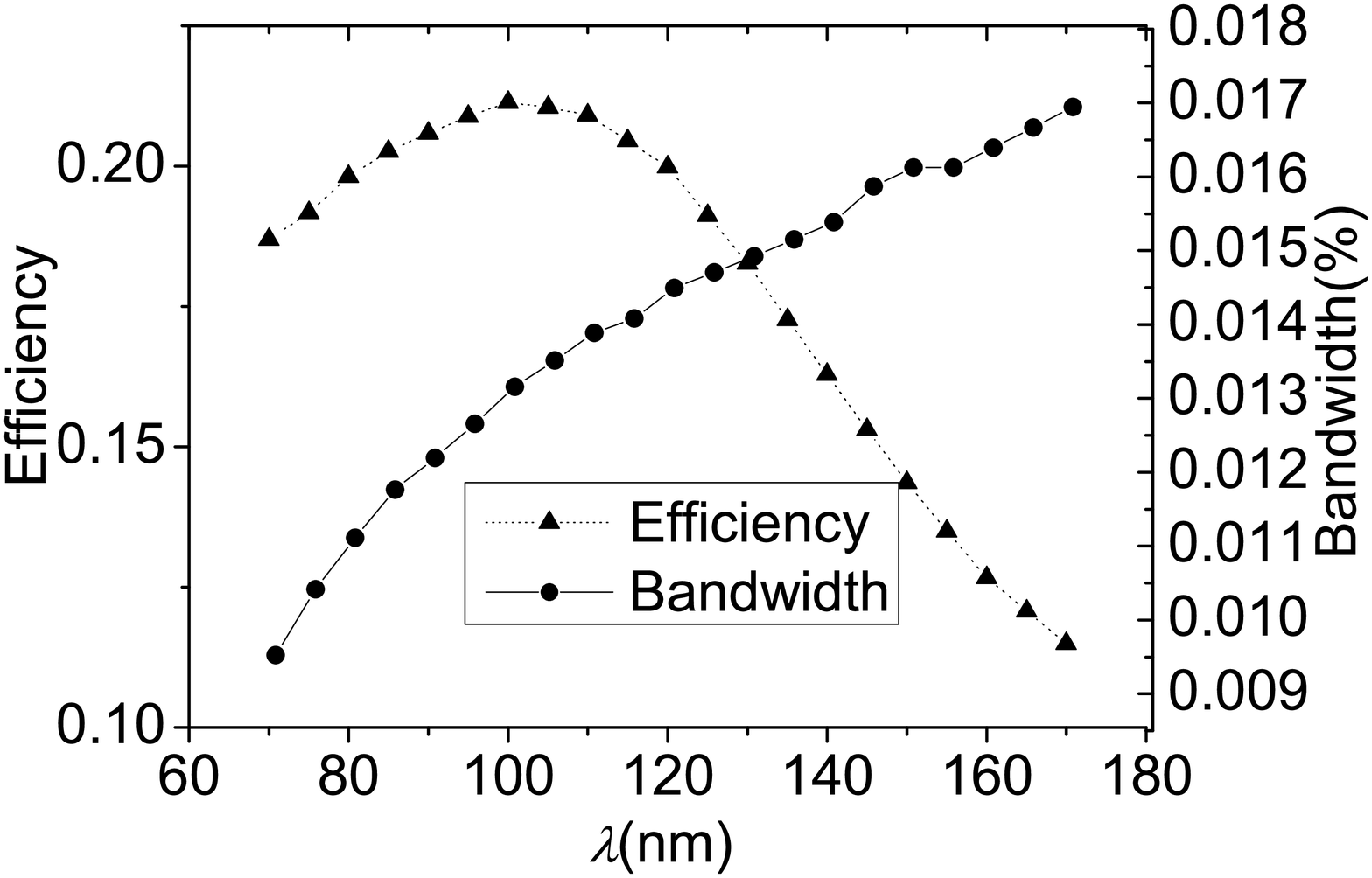}
\figcaption{\label{fig4} The transport efficiency and bandwidth of the monochromator.}
\end{center}

\begin{center}
\includegraphics[width=4.0cm]{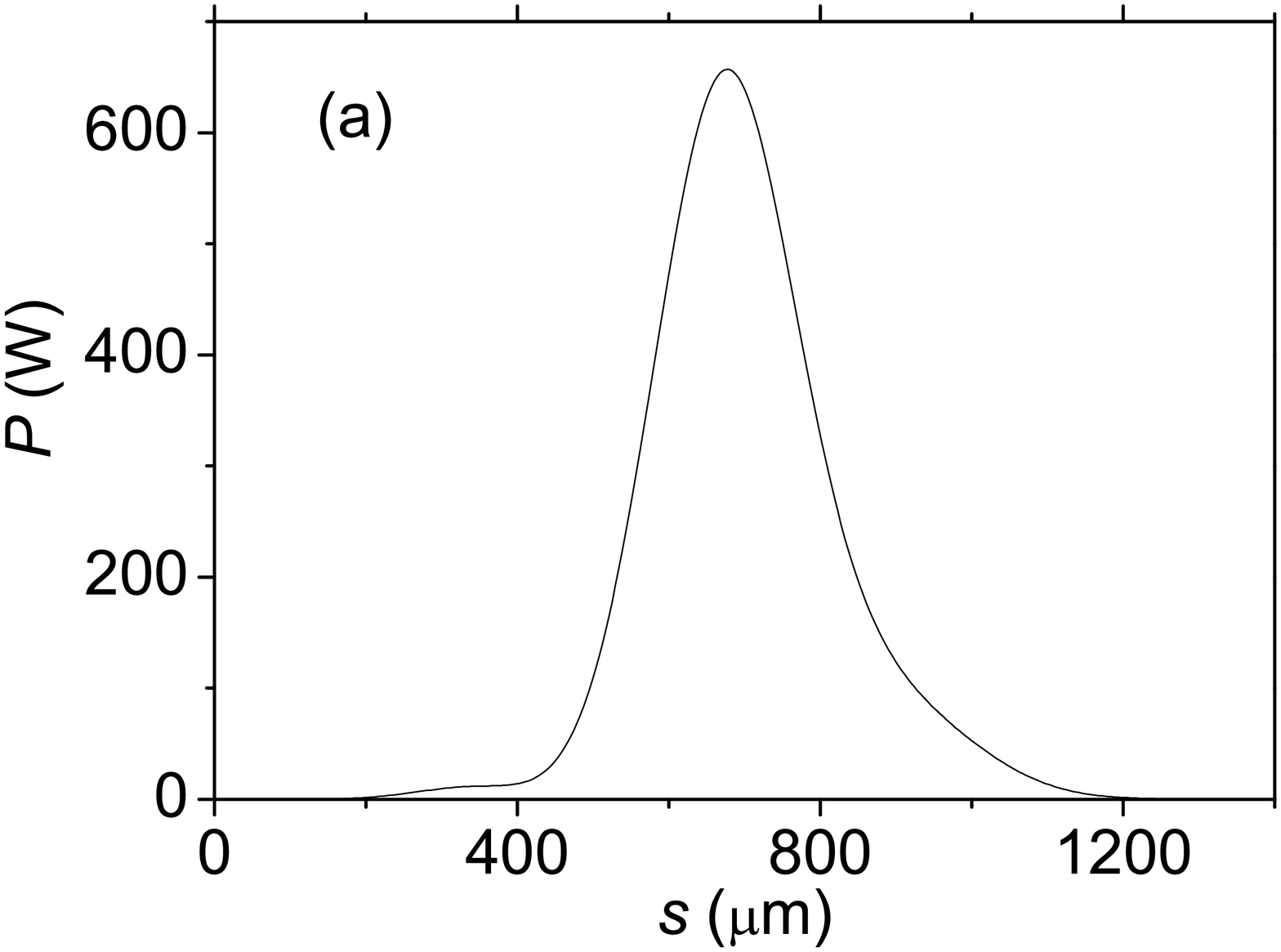}
\includegraphics[width=4.0cm]{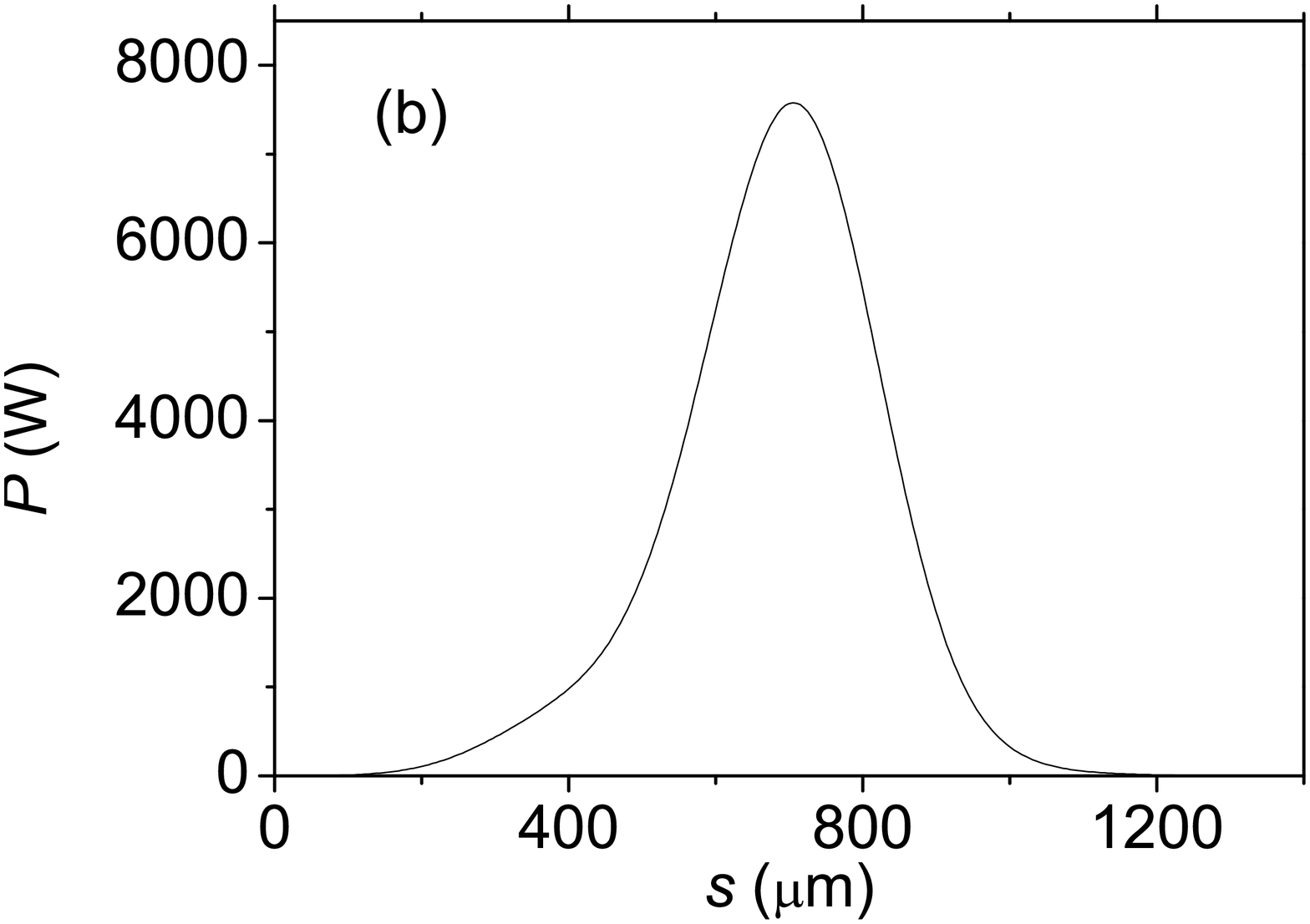}
\figcaption{\label{fig4} The seed pulse for the second section at (a) 70 nm and (b) 170 nm.}
\end{center}
To determine the length of the first section, firstly we estimate the shot noise power at the beginning of the second section, which is in the order of several to tens Watts. Another important point is the design of the monochromator, which usually consists of a grating [3]. The monochromator in this proposal is composed of three mirrors and a rotational grating traditionally. The transport efficiency and bandwidth are given in Fig.4. It can be found that the transport efficiency at short wavelength is higher than that at long wavelength, which can compensate the seed power in shorter wavelength to a certain degree. Based on these, the undulator length of the first section is selected to be 6.4 m. In this case, the time structure of the seed pulse for the second section at 70 nm and 170 nm is given in Fig.5 by time-dependent simulation. Obviously, the power is high enough to seed the second section. we should indicate that we scan the wavelength in the nearby area when the radiation pulse generated in the first section passes through the monochromator, because the radiation at the target wavelength maybe locate in the trough between two spikes in the spectrum. However, in the experiment, it is impossible to scan the wavelength for every coming pulse. For this reason the output radiation of the second section will fluctuate, which is an inherent disadvantage of self-seeding scheme as the SASE process starts up from shot-noise.

In the second section, it remains that the power gain length has a big variation range so that the saturation length is changed with the radiation wavelength. Thus, the undulator length is fixed by the saturation length of 70 nm radiation, which is about seven undulators. However, in the long wavelength range, the undulator length is longer than the saturation length. Then tapering technique is considered to enhance the radiation power  in long wavelength range [11-13]. In detail, the tapering starts from the undulator most close to saturation point, and the tapering amplitude is achieved by rough calculation first and then numerical scan. The radiation properties at several typical wavelengths is shown in Table 3. The pulse energy is basically stable when tapering does not open. When tapering is applied, the pulse energy of the long wavelength increases by several times. Fortunately, it meets the demand of the FEL users perfectly. We also compare the spectrum with and without tapering, as Fig.5 shows the 170 nm case.
\begin{center}
\includegraphics[width=8.0cm]{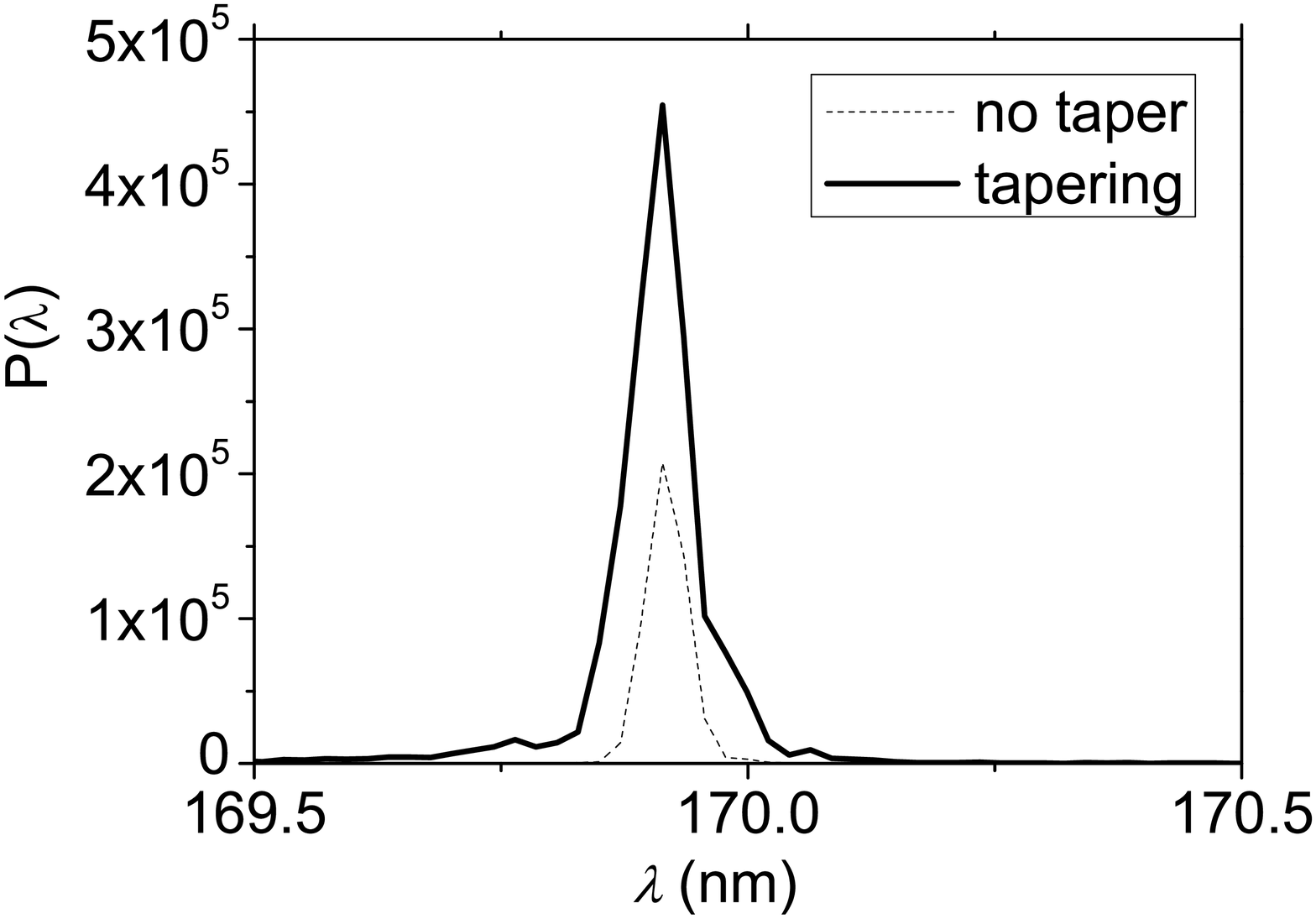}
\figcaption{\label{fig6} The spectrum for 170 nm radiation at saturation point without taper (dashed) and at the end of the second section with taper (line).}
\end{center}

\end{multicols}

\begin{center}
\tabcaption{ \label{tab3}  The radiation properties of several typical wavelengths.}
\footnotesize
\begin{tabular*}{170mm}{@{\extracolsep{\fill}}ccccccc}
\toprule $\lambda _s$ /nm & $L_{sat} $ /m  & Tapering start point /m & $\delta a_u $ & $\emph{E}_{pulse}$ (no taper) /mJ &  $\emph{E}_{pulse}$ (tapered) /mJ \\
\hline
70 & 12.352 & No & $-$ & 0.245 & $-$\\
85 & 12.352 & No & $-$ & 0.274 & $-$\\
115 & 10.752 & 9.152 & $1.5\%$ & 0.316 & 1.68\\
150 & 8.96 & 9.152\hphantom{0} & $2\%$& 0.337 & 2.14\\
170 & 8.352 & 7.328\hphantom{0} & $3\%$& 0.564 & 2.89\\
\bottomrule
\end{tabular*}%
\end{center}

\begin{multicols}{2}

\section{Summary}
We have preliminarily designed a wavelength tunable VUV source using self-seeding scheme. In this paper, emphasis was put on the global physical design but not the detail cell design. To reach the goal of the broad radiation wavelength range (70-170 nm), we divide the target wavelength range into three subareas, and in each subarea a constant electron energy is used and the transverse focusing system is fixed. Higher energy is used for shorter wavelength to increase the beam power and compensate the longer gain length and lower FEL efficiency. The simulation results shows that the scheme of dividing into three subareas has an acceptable performance. But if more subareas be divided or the difference of electron energy of the each subarea be increased, maybe a better performance can be achieved.

More work has to be done for the regular design, such as the bypass chicane design, the studies on tolerance and stability, and so on.

\end{multicols}

\vspace{-1mm}
\centerline{\rule{80mm}{0.1pt}}
\vspace{2mm}

\begin{multicols}{2}

\end{multicols}
\clearpage


\begin{thebibliography}{90}

\vspace{3mm}

\bibitem{lab1} Technical Design Report of DCLS, March 2012  (in chinese)

\bibitem{lab2} Yu L H. Phys. Rev. A, 1991, \textbf{44}: 5178

\bibitem{lab3} J. Feldhaus et al., Optics. Comm., 1997, \textbf{140}: 341

\bibitem{lab4} E. Saldin et al.,  Nucl. Instrum. Methods. A, 2001, \textbf{475}: 357

\bibitem{lab5} E. Saldin et al., Nucl. Instrum. Methods. A, 2000,  \textbf{445}: 178

\bibitem{lab6} Y. Ding et al., Phys. Rev. ST ST Accel. Beams, 2010, \textbf{13}: 060703

\bibitem{lab7} J. Wu et al., Staged self-seeding scheme for narrow bandwidth , ultrashort X-ray harmonic generation free electron laser at LCLS, Proc. of FEL Conference, Malmo, Sweden: 2010. 266

\bibitem{lab8} J. Amann et al., Nature photonics, 2012, \textbf{6}: 693

\bibitem{lab9} Murphy J et al. Opt. Commun, 1985, \textbf{53}: 197

\bibitem{lab10} Reiche S. Nucl. Instrum. Methods. A, 1999, \textbf{429}: 243

\bibitem{lab11} A. Lin and J.M. Dawson, Phys. Rev. Lett. 1979, \textbf{42}: 1670

\bibitem{lab12} D. A. Jaroszynski et al., Phys. Rev. Lett., 1995, \textbf{74}: 2224

\bibitem{lab13} W. M. Fawley et al., Nucl. Instr. and Meth. A, 2002, \textbf{483}: 537
\end{thebibliography}
\end{document}